\documentclass{article}
\usepackage{array,psfig,times,url,epsf,epsfig}
\begin{document}
\title{Public-Key Cryptography Standards: PKCS}
\author{Yongge Wang, Ph.D., University of North Carolina at Charlotte}

\date{}
%\author{}
\maketitle
{\small
Outline: 
\ref{Intro} Introduction, 
\ref{pkcs1} PKCS \#1: RSA Cryptography Standard, 
\ref{pkcs3} PKCS \#3: Diffie-Hellman Key Agreement Standard (Outdated),
\ref{pkcs5} PKCS \#5: Password-Based Cryptography Standard,
\ref{pkcs6} PKCS \#6: Extended-Certificate Syntax Standard (Historic),
\ref{pkcs7} PKCS \#7 and RFC 3369: Cryptographic Message Syntax (CMS),
\ref{pkcs8} PKCS \#8: Private-Key Information Syntax Standard,
\ref{pkcs9} PKCS \#9: Selected Object Classes and Attribute Types,
\ref{pkcs10} PKCS \#10: Certification Request Syntax Standard,
\ref{pkcs11} PKCS \#11: Cryptographic Token Interface Standard,
\ref{pkcs12} PKCS \#12: Personal Information Exchange Syntax Standard,
\ref{pkcs15} PKCS \#15: Cryptographic Token Information Syntax Standard,
\ref{conclusion} An Example.}

{\bf Key works}. ASN.1, public key cryptography, digital signature, 
encryption, key establishment scheme,
public key certificate, cryptographic message syntax, 
cryptographic token interface (cryptoki).
\begin{abstract}
Cryptographic standards serve two important goals:
making different implementations interoperable and avoiding various
known pitfalls in commonly used schemes.
This chapter discusses Public-Key Cryptography Standards  (PKCS)
which have significant impact on the use of public key cryptography
in practice. PKCS standards are a set of standards,
called PKCS \#1 through \#15. 
These standards cover RSA encryption, RSA signature,
password-based encryption, cryptographic message syntax, 
private-key information syntax, selected object classes and attribute 
types, certification request syntax,
cryptographic token interface, personal information exchange
syntax, and cryptographic token information syntax.
The PKCS standards are published by RSA Laboratories.
Though RSA Laboratories solicits public opinions and advice for 
PKCS standards, RSA Laboratories retain sole decision-making 
authority on all aspects of PKCS standards. PKCS has been the
basis for many other standards such as S/MIME.
\end{abstract}

\section{Introduction}
\label{Intro}
{\em Public key cryptography} is based on
asymmetric cryptographic algorithms that use two related keys,
a {\em public key} and a {\em private key}; the two keys have the property
that, given the public key, it is computationally infeasible
to derive the private key. A user publishes his/her public key in a 
public directory such as an LDAP directory and keeps his/her private
key to himself/herself. 

According to the purpose of the algorithm, there are public-key
encryption/decryption algorithms and signature algorithms.
An encryption algorithm could be used to encrypt a data (for example,
a symmetric key) using the public key so that only the recipient 
who has the corresponding private key could decrypt the data.
Typical public key encryption algorithms are RSA and ECIES 
(Elliptic Curve Integrated Encryption Scheme, see, SECG 2000).
A signature algorithm together with a message digest algorithm 
could be used to transform a message of any length
using the private key to a {\em signature} in such a way that, without the
knowledge of the private key, it is computationally infeasible 
to find two messages with the same signature, to find a message for a 
pre-determined signature, or to find a signature for a given message.
Anyone who has the corresponding public key could verify the validity of the
signature. Typical public key digital signature algorithms are 
RSA, DSA, and ECDSA.

There have been extensive standardization efforts for 
public key cryptographic techniques. The major standards organizations that 
have been involved in public key cryptographic techniques are:
\begin{itemize}
\item ISO/IEC. The International Organization for Standardization (ISO)
and the International Electrotechnical Commission (IEC) 
(individually and jointly) have been developing a series of standards 
for application-independent cryptographic techniques. ISO has 
also been developing
bank security standards under the ISO technical committee TC86--Banking
and Related Financial Services.
\item ANSI. The American National Standards Institute (ANSI) have been
developing public key cryptographic technique standards for financial
services under Accredited Standards Committee (ASC) X9. For example,
they have developed the standards ANSI X9.42 (key management
using Diffie-Hellman), ANSI X9.44 (key establishment using factoring-based
public key cryptography), and ANSI X9.63
(key agreement and key management using ECC).
\item NIST. The National Institute of Standards and Technology
(NIST) has been developing public key cryptography standards for use
by US federal government departments. These standards are released
in Federal Information Processing Standards (FIPS) publications.
\item IETF. The Internet Engineering Task Force has been developing
public key cryptography standards for use by the Internet community.
These standards are published in Requests for Comments (RFCs).
\item IEEE. The IEEE 1363 working group has been publishing standards
for public key cryptography, including IEEE 1363-2000, IEEE 1363a,
IEEE P1363.1, and IEEE P1363.2.
\item Vendor-specific standards. This category includes PKCS standards
that we will describe, SEC standards, and others. 
Standards for Efficient Cryptography (SEC) \#1 and \#2 are 
elliptic curve public key cryptography standards that have been developed 
by Certicom Corp.~in cooperation with secure systems developers world-wide. 
\end{itemize}
The PKCS standards, developed by RSA Laboratories 
(a Division of RSA Data Security Inc.) in cooperation with secure 
systems developers worldwide for the purpose of accelerating the 
deployment of public-key cryptography, are widely implemented in practice, 
and periodically updated. Contributions from the PKCS standards
have become part of many formal and de facto standards, including
ANSI X9 documents, IETF documents, 
and SSL/TLS (Secure Socket Layer/Transport Layer Security).
The parts and status of PKCS standards are listed in Table \ref{pkcstable}
and are discussed in details in the following sections.
The descriptions are largely adapted from the PKCS documents themselves.
In Section \ref{conclusion}, we give an example application which
uses all these PKCS standards.
{\small \begin{center}
\begin{table}[htb]
\caption{PKCS Specifications}
\label{pkcstable}
\begin{center}
\begin{tabular}{|c|l|l|} \hline
{\bf No.} & {\bf PKCS title} & {\bf Comments}\\ \hline
1 & RSA Cryptography Standard & \\ \hline
2,4 & & incorporated into  PKCS \#1\\ \hline
3 & Diffie-Hellman Key Agreement Standard & superseded by 
        IEEE 1363a etc.\\ \hline
5 & Password-Based Cryptography Standard & \\ \hline
6 & Extended-Certificate Syntax Standard & never adopted \\ \hline 
7 & Cryptographic Message Syntax Standard & superseded by 
        RFC 3369 (CMS) \\ \hline
8 & Private-Key Information Syntax Standard & \\ \hline
9 & Selected Object Classes and Attribute Types & \\ \hline
10& Certification Request Syntax Standard & \\ \hline
11& Cryptographic Token Interface Standard & referred to as CRYPTOKI\\ \hline
12& Personal Information Exchange Syntax Standard& \\ \hline
13& ({\em reserved for ECC}) & never been published\\ \hline
14& ({\em reserved for pseudo random number generation}) & 
        never been published\\ \hline
15& Cryptographic Token Information Syntax Standard& \\ \hline
\end{tabular}
\end{center}
\end{table}
\end{center}
}
%PKCS \#3 is superseded by IEEE 1363a (2003), ANSI X9.42, ANSI X9.44, and
%ANSI X9.63. PKCS \#6 has never been adopted. PKCS \#7 has been superseded by 
%IETF Cryptographic Message Syntax (CMS), which is used in S/MIME
%and other applications.

\section{PKCS \#1: RSA Cryptography Standard}
\label{pkcs1}
PKCS \#1 v2.1 provides standards for implementing RSA algorithm-based
public key cryptographic encryption schemes and digital signature 
schemes with appendix. It also defines corresponding 
ASN.1 syntax for representing keys and for identifying the schemes.

RSA is a public-key algorithm invented by Rivest, Shamir, and Adleman (1978)
which is based on the exponentiation modulo the 
product of two large prime numbers. The security of RSA algorithm 
is believed to be based on the hardness of factoring the product of 
large prime numbers. In PKCS \#1 v2.1, multiprime RSA scheme is introduced. 
Multiprime RSA means that the modulus isn't
the product of two primes but of more than two primes.
This is used to increase performance of RSA cryptographic primitives.
In particular, in multiprocessor environments, one can
exponentiate modulo each prime and then apply the Chinese remainder theorem 
to get the final results. However, one should be aware that the security
strength of multiprime RSA is a little different from the original
RSA scheme. If we assume that the best way to attack multiprime RSA is to 
factorize the modulus and the best factorization algorithm is
the Number Field Sieve (NFS) algorithm, then we can compute 
the approximate strength of some multiprime RSA schemes as listed in 
Table \ref{stable}, where $u$ is the number of primes.
Similar tables for two primes RSA could be found in literatures,
e.g., Lenstra and Verheul (2001).
\begin{center}
\begin{table}[htb]
\caption{Security strength of multiprime RSA schemes}
\label{stable}
\begin{center}
\begin{tabular}{|c|c|c||c|c|c|} \hline
Symmetric Key Size & RSA Modulus Size & $u$&Symmetric Key Size& RSA 
Modulus Size & $u$ \\ \hline
{\bf 80} & 1024 & 2 & {\bf 192} & 7680 & 4   \\ \hline
73 & 1024 & 3 & 175 & 7680 & 5 \\ \hline
{\bf 112} & 2335 & 3 &158 & 7680 & 6 \\ \hline
100 & 2335 & 4& 144 & 7680 & 7\\ \hline
88 & 2335 & 5& 125 & 7680 & 9\\ \hline
{\bf 128} & 3072 & 3& {\bf 256} & 15360 & 5\\ \hline
117 & 3072 & 4&235 & 15360 & 6\\ \hline
103 & 3072 & 5& 215 & 15360 & 7\\ \hline
93 & 3072 & 6& 199 & 15360 & 8\\ \hline
\end{tabular}
\end{center}
\end{table}
\end{center}

\subsection{RSA keys}
Let $n=r_1\cdots r_{u}$ be the product 
of $u\ge 2$ distinct prime numbers of approximately 
the same size ($|n|/u$ bits each), where $|n|$ denotes the number of 
bits in $n$. For the case of $u=2$, 
one normally uses $p$ and $q$ to denote the two prime numbers,
that is, $n=pq$. A typical size for $n$ is 
1024 bits, and $u=2$. Let $e,d$ be two integers satisfying 
$e\cdot d \equiv 1 \ (\mbox{mod }\chi (n))$, where $\chi(n)$ is the least
common multiple of $r_1-1, r_2-1, \ldots, r_u-1$. 
We call $n$ the RSA {\em modulus}, $e$ the {\em encryption exponent},
and $d$ the {\em decryption exponent}. The pair $(n,e)$ is the 
{\em public key} and the pair $(n,d)$ is called the {\em secret key}
or {\em private key}. The public key is public and one can use it
to encrypt messages or to verify digital signatures. The private key 
is known only to the owner of the private key and can be
used to decrypt ciphertexts or to digitally sign messages.

In order to efficiently decrypt ciphertexts and to efficiently generate
digital signatures, the private key may include further information 
such as the first two prime factors and  CRT exponents and CRT 
coefficients of each prime factor. For a prime factor $r_i$, its CRT exponent
is a number $d_i$ satisfying $e\cdot d_i\equiv 1\ (\mbox{mod }(r_i-1))$,
and its  CRT coefficient $t_i$ is a positive integer 
less than $r_i$ satisfying $R_i\cdot t_i\equiv 1\ (\mbox{mod }r_i)$,
where $R_i = r_1\cdot r_2\cdot\ldots\cdot r_{i-1}$.
PKCS \#1 v2.1 specifies the format
for such kind of enhanced private keys.

\subsection{RSA encryption schemes}
We begin by describing a basic version of 
RSA encryption scheme. A message is an integer $m<n$. To encrypt $m$, 
one computes $c\equiv m^e\mbox{ mod }n$.
To decrypt the ciphertext $c$, the legitimate receiver computes
$c^d\mbox{ mod }n$. Indeed, 
$$c^d\equiv m^{ed}\equiv m\mbox{ mod }n,$$ where
the last equality follows by Euler's theorem.

For performance reasons, RSA is generally not used to encrypt 
long data messages directly. Typically, 
RSA is used to encrypt a secret key and the data is encrypted with 
the secret key using a secret key cryptography scheme such as DES or AES.
Thus the actual data to be encrypted by RSA scheme is generally much 
smaller than the modulus and the message (secret key) needs to be 
padded to the same length of the modulus before encryption. 
For example, if AES-128 is used, then an AES key is 128 bits.
Another reason for a standardized padding
prior to encryption using some randomness is that the 
basic version of RSA encryption scheme is not secure and is 
vulnerable to many attacks. PKCS \#1 v2.1 provides two message padding 
methods: EME-PKCS1-v1\_5 and EME-OAEP. 

\subsubsection{RSAES-PKCS1-v1\_5 padding}
After EME-PKCS1-v1\_5 padding to $M$, the padded message $EM$ looks 
as follows:
\begin{center}
$EM=\ \ $\begin{tabular}{|c|c|c|c|c|}\hline
0x00 & 0x02 & random octets & 0x00 & $M$\\ \hline
\end{tabular}
\end{center}
where ``random octets'' consists of pseudo-randomly generated 
nonzero octets and $0\mbox{x}00$ octet is used to delimit 
the padding from the actual data.
The length of ``random octets'' is at least eight octets.
The top octet 0x00 guarantees that the padded message 
is smaller than the modulus $n$ (PKCS \#1 v2.1 specifies that 
the high-order octet of the modulus must be non-zero). 
If the padded message $EM$ were 
larger than $n$,  decryption would produce $EM$ mod $n$ instead of $EM$. 
The next octet $0\mbox{x}02$ is the format type.
The value  $0\mbox{x}02$ is used to encryption and the value
$0\mbox{x}01$ is used for signature padding format  RSASSA-PKCS1-v1\_5 
(RSASSA-PKCS1-v1\_5 is no long recommended by RSA Lab.).
The resulting padded message $EM$ is
$|n|$ bits and is directly encrypted using the basic version of
RSA.

Bleichenbacher (1998) pointed out that improper implementation
of the above padding method can lead to disastrous consequences.
When the encrypted message arrives at the receiver's computer,
an application decrypts it, checks the initial block, and strips off the 
random pad. However, some applications check for the two initial
blocks 0x00 02 and if it is incorrect, they send the error
message saying ``invalid ciphertext''. These error messages can help
the attacker to decrypt ciphertext of his choice.
PKCS \#1 v2.1 recommends certain easily implemented countermeasures 
to thwart this attack. Typical examples include the addition
of structure to the data to be encoded, rigorous checking of 
PKCS \#1 v1.5 conformance in decrypted messages, and the consolidation of 
error messages in a client-server protocol based on PKCS \#1 v1.5.

\subsubsection{RSAES-OAEP padding}
EME-OAEP is based on Bellare and Rogaway's (1995) Optimal Asymmetric 
Encryption scheme.
Assuming that it is difficult to inverse the RSA function and the 
mask generation function in the OAEP padding has appropriate 
properties, RSAES-OAEP is proven to be secure in a stronger sense.
The reader is referred to Bellare and Rogaway (1995) for 
details.

Let $k$ be the length in octets of the recipient's RSA modulus,
$k_0<k$ be an integer, $H$ be a hash function whose outputs are
$k_0$-octets, and MGF be the mask generation function. For 
an input octet string $x$ and an integer $i$, $\mbox{MGF}(x, i)$ 
outputs a string of $i$ octets. Let $M$ be the $k_1$-octets message
such that $k_1<k - 2k_0 - 2$, 
and $L$ be an optional label (could be an empty string) to be associated 
with the message. EME-OAEP first converts the message $M$ to 
a $(k-k_0-1)$-octets data block $DB$ that looks as follows:
\begin{center}
$DB=\ \ $\begin{tabular}{|c|c|c|c|}\hline
$H(L)$ & random octets & 0x01 & $M$\\ \hline
\end{tabular}
\end{center}
where ``random octets'' consists of pseudo-randomly generated octets.
The length of ``random octets'' could be zero. EME-OAEP then chooses
a random $k_0$-octets string $r$, and generates the OAEP padded message
$EM$ as follows:
\begin{center}
$EM=\ \ $\begin{tabular}{|c|c|c|}\hline
0x00 & $r\oplus \mbox{MGF}(DB\oplus\mbox{MGF}(r,k-k_0-1) ,k_0)$ 
&  $DB\oplus\mbox{MGF}(r,k-k_0-1)$  \\ \hline
\end{tabular}
\end{center}
The resulting padded message $EM$ is $k$-octets and is directly 
encrypted using the basic version of
RSA. For decryption operations, EME-OAEP decoding method could 
be constructed directly.

\subsection{RSA signature schemes with appendix}
We begin by describing a basic version of 
RSA signature scheme with appendix. A message is an integer $m<n$. 
To sign $m$, the owner of the 
private key $(n,d)$ computes the signature $s\equiv m^d\mbox{ mod }n$.
To verify that $s$ is a signature on $m$ from the legitimate 
owner of the private key $(n,d)$, one uses the corresponding
public key $(n,e)$ to compute $m'\equiv s^e\ (\mbox{mod }n)$.
If $m'=m$, then the signature is valid, otherwise, the signature is invalid.

The basic version of RSA signature scheme 
can only generate signatures on messages less than $|n|$ bits.
In addition, the basic version of RSA signature scheme is not secure.
To address these issues, 
in practice, one first computes a message digest from 
a given message using a hash function such as MD5 or SHA-1.
The message digest is encoded using an encoding method
and the resulting string is converted to an integer and is
supplied to the basic RSA signature primitive.

PKCS \#1 v2.1 provides two encoding methods
for encoding message digests: EMSA-PKCS1-v1\_5 encoding
and EMSA-PSS encoding. Correspondingly there are 
two signature schemes with appendix: 
RSASSA-PSS and RSASSA-PKCS1-v1\_5. Although no attacks are known 
against RSASSA-PKCS1-v1\_5, in the interest of increased robustness, 
RSASSA-PSS is recommended for eventual adoption in new applications. 
RSASSA-PKCS1-v1\_5 is included in PKCS \#1 v2.1 for compatibility 
with existing applications and we will not discuss it here.
EMSA-PSS is based on the work of Bellare and Rogaway's (1996).
Assuming that computing $e$th roots modulo $n$ is infeasible and 
the hash and mask generation functions in EMSA-PSS have appropriate 
properties, RSASSA-PSS provides secure signatures. This assurance 
is provable in the sense that the difficulty of forging signatures 
can be directly related to the difficulty of inverting the RSA 
function, provided that the hash and mask generation functions 
are viewed as black boxes or random oracles. The reader is referred to
Bellare and Rogaway's (1996) for more details.

Let $k$ be the length in octets of the RSA modulus,
$H$ be a hash function whose outputs are
$k_0$ octets ($k_0<k$), and MGF be the mask generation function. For 
an input octet string $x$ and an integer $i$, $\mbox{MGF}(x, i)$ 
outputs a string of $i$ octets. Let $M$ be the message to be signed.
EMSA-PSS first constructs octet strings $M'$ and $DB$ as follows:
$$
M' \ = \  \begin{tabular}{|c|c|c|}\hline
0x00 00 00 00 00 00 00 00 & \mbox{$H(M)$} & salt\\ \hline
\end{tabular}\ ,
\quad \quad 
DB\ = \  \begin{tabular}{|c|c|c|}\hline
PS & 0x01 & salt\\ \hline
\end{tabular}
$$
where ``salt'' and ``PS'' consist of pseudo-randomly generated octets.
The lengths of ``salt'' and ``PS''  could be zero, and the length of 
$DB$ is $k-k_0-1$ octets.

EMSA-PSS then constructs the octet string $EM'$ as follows:
\begin{center}
$EM'=\ \ $\begin{tabular}{|c|c|c|}\hline
$DB\oplus\mbox{MGF}(H(M'),k-k_0-1)$ & $H(M')$ & 0xbc \\ \hline
\end{tabular}
\end{center}
Assume that the RSA modulus has $|n|$ bits, then the encoded
string $EM$ is obtained by setting the leftmost
$8k-|n|+1$ bits of the leftmost octet in $EM'$ to zero.
The resulting encoded string $EM$ is $k$ octets and is directly 
signed using the basic version of
RSA signature scheme. 
The EMSA-PSS decoding process could to be constructed
directly.

\section{PKCS \#3: Diffie-Hellman Key Agreement Standard (Outdated)}
\label{pkcs3}
PKCS \#3 v1.4 describes a method for implementing Diffie-Hellman 
key agreement, whereby two parties can agree upon a secret key 
that is known only to them. 
PKCS \#3 is superseded by modern treatment of key establishment 
schemes specified in IEEE 1363a (2003), ANSI 9.42, ANSI X9.44, and
ANSI X9.63 etc. Basically there are two types of key establishment
schemes:
\begin{enumerate}
\item Key agreement scheme: a key establishment scheme in which the keying
data established is a function of contributions provided by both entities in 
such a way that neither party can predetermine the value of the keying data.
Diffie-Hellman key agreement scheme is an example of this category.
%assume that $p$ is a large prime, 
%in a Diffie-Hellman key agreement scheme, two parties choose
%random $x_1$ and $x_2$ respectively, exchanges $g^{x_1}$ 
%and $g^{x_2}$, and the shared keying material is $g^{x_1x_2}$.
\item Key transport scheme: a key establishment scheme in which the keying
data established is determined entirely by one entity. For example, one party
chooses a random session key, encrypts it with the other party's public key,
and sends the encrypted session key to the other party. The other party 
can then decrypt the session key. A special case of key transport scheme is 
the key wrap scheme in which the session key is encrypted with a pre-shared 
secret using a secret key cipher such as DES or AES.
\end{enumerate}

\section{PKCS \#5: Password-Based Cryptography Standard}
\label{pkcs5}
In many applications of public-key cryptography, user security is 
ultimately dependent on one or more secret text values or passwords.
For example, user's private key is usually encrypted with a password
and the encrypted private key is kept in storage devices 
(see Section \ref{pkcs8}).
However, there are two essential problems regarding to password application:
(1) A password is not directly applicable as a key to any conventional 
cryptosystem; (2) Passwords are often chosen from a relatively 
small space. Thus special care is required to defend against search attacks.
PKCS \#5 provides a general mechanism to achieve an enhanced security
for password-based cryptographic primitives, covering key derivation 
functions, encryption schemes, message-authentication schemes, and 
ASN.1 syntax identifying the techniques. It should be noted that 
other password based cryptographic techniques are currently 
under standardization process in IEEE 1363.2.

\subsection{Key derivation functions}
A password-based key derivation function
produces a key from a password, a random salt value, and an iteration count.
The salt is not secret and serves the purpose of producing a 
large set of keys for one 
given password, among which one is selected at random according to the salt. 
An iteration count serves the purpose of increasing the cost 
of producing keys from a password, thereby also increasing 
the difficulty of attack. 
PKCS \#5 v2.0 specifies two password-based key derivation functions 
PBKDF1 and PBKDF2. PBKDF1 is included in PKCS \#5 v2.0  only for 
compatibility with existing applications following PKCS \#5 v1.5, 
and is not recommended for new applications.

PBKDF2 applies a pseudorandom function to derive keys. 
The length of the derived key is essentially unbounded. 
However, the maximum length for the derived key 
may be limited by the structure of the underlying pseudorandom function. 
Let  $H$ be a pseudorandom function whose outputs are
$hLen$  octets, $dkLen\le (2^{32} - 1)\times hLen$ 
be the intended length in octets 
for the derived key, $P$ be the password (an octet string), 
$S$ be an eight-octet salt string, and $c$ be an iterating count.
For each integer $i$, by repeatedly hashing the password, salts, etc.,
one gets a sequence of $hLen$-octets strings:
$$U_1^{i} =  H(P, S ||\mbox{INT}(i)),\ 
U_2^{i} = H(P, U_1^{i}),\ 
\ldots,\  U_c^{i} = H(P, U_{c-1}^{i}),$$
where $\mbox{INT}(i)$ is a four-octet encoding of the integer $i$, 
most significant octet first. Then one computes the  $hLen$-octet strings
$T_i = U_1^{i} \oplus U_2^{i} \oplus\ldots\oplus U_c^{i}$
for each $i$. The derived key is the first  $dkLen$-octet 
of the string $T_1||T_2||T_3||\cdots$. In another word, 
let $l =\lceil dkLen / hLen\rceil$ be the number of $hLen$-octet blocks
in the derived key, rounding up, and $r = dkLen - (l - 1)\times hLen$ be 
the number of octets in the last block.
Then the $dkLen$-octet derived key $DK=\mbox{PBKDF2}(P, S, c, dkLen)$ 
looks as follows:
\begin{center}
$DK=$ \begin{tabular}{|c|c|c|c|}\hline
$T_1$ & $T_2$ & $\cdots$ & $T_l[0..r-1]$\\ \hline
\end{tabular}
\end{center}

\subsection{Encryption schemes}
PKCS \#5 v2.0 specifies two encryption schemes PBES1 and PBES2.
PBES1 is included in PKCS \#5 v2.0 only 
for compatibility with PKCS \#5 v1.5, and is not recommended 
for new applications.
PBES2 combines the password-based key derivation function PBKDF2 
with an underlying encryption scheme ${\cal E}$. Let $M$ be the 
message to be encrypted, $P$ be the password, $k$ be the key length
in octets for ${\cal E}$. For the PBES2 encryption, one first
selects a salt $S$ and an iteration count $c$, then one 
computes the derived $k$ octets key $DK = \mbox{PBKDF2}(P, S, c, k)$.
The ciphertext $C$ for $M$ is: $C={\cal E}_{DK}(M)$.
The decryption operation for PBES2 can be done similarly.

\subsection{Message authentication schemes}
In a {\em password-based message authentication scheme}, the
MAC generation operation produces a message authentication code
from a message under a password, and the MAC verification
operation verifies the message authentication 
code under the same password. PKCS \#5 v2.0 defines the password-based 
message authentication scheme PBMAC1 which combines the password-based 
key derivation function PBKDF2 with an underlying message authentication 
scheme ${\cal A}$. 

Let $M$ be the message to be authenticated, $P$ be the password, 
$k$ be the key length
in octets for ${\cal A}$. For PBMAC1, one first
selects a salt $S$ and an iteration count $c$, then one 
computes the derived $k$ octets key $DK = \mbox{PBKDF2}(P, S, c, k)$.
The message authentication code $T$ can be computed as
$T={\cal A}(M,DK)$. 
The MAC verification operation for PBMAC1 can be done similarly.

\section{PKCS \#6: Extended-Certificate Syntax Standard (Historic)} 
\label{pkcs6}
When PKCS \#6 was drafted, X.509 was in version 1.0
and no {\tt extensions} component was defined in the certificate. 
An X.509 v3 certificate can contain information about 
a given entity in the {\tt extensions} component.
Since the introduction of X.509 v3, the status of PKCS \#6 is historic.

\section{PKCS \#7 and RFC 3369: Cryptographic Message Syntax (CMS)}
\label{pkcs7}
PKCS \#7 has been superseded by IETF RFC 3369 (Housley 2002): 
cryptographic message syntax (CMS), which is the basis for the 
S/MIME specification. CMS defines the syntax that is used to
digitally sign, digest, authenticate, or encrypt arbitrary message content.
In particular, CMS describes an encapsulation syntax for data protection. 
The syntax allows multiple encapsulations; one encapsulation envelope 
can be nested inside another. Likewise, one party can digitally sign some
previously encapsulated data. In the CMS syntax, arbitrary attributes,
such as signing time, can be signed along with the message content,
and other attributes, such as countersignatures, can be
associated with a signature. A variety of architectures for 
certificate-based key management (e.g., the one defined
by the IETF PKIX working group) are supported in CMS.

The CMS values are generated using ASN.1 with BER-encoding and are 
typically represented as octet strings. When transmitting 
CMS values in systems (e.g., email systems) that do not support 
reliable octet strings transmission, one should use additional
encoding mechanisms that are not addressed in CMS.

\begin{center}
\begin{figure}[htb]
\centering{\includegraphics{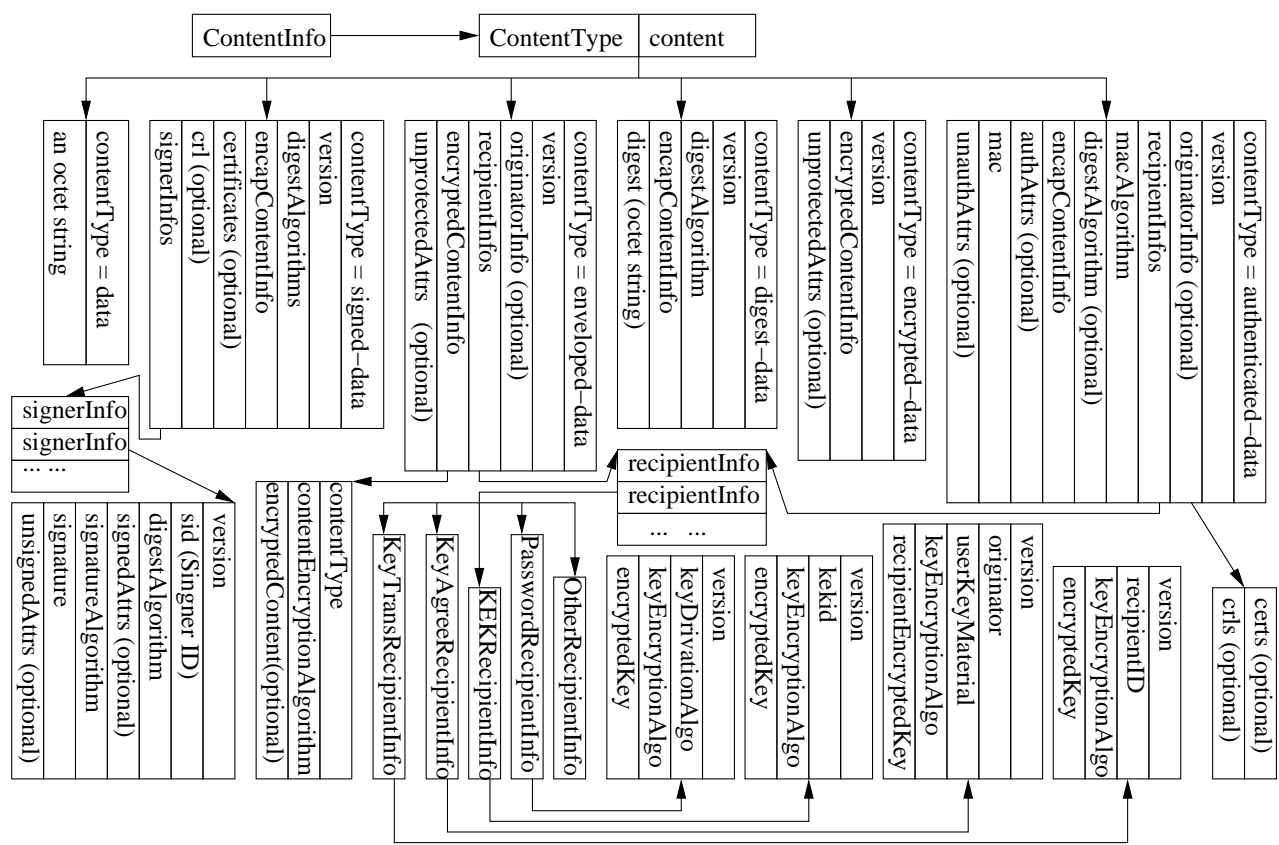}}
\caption{CMS content types and their fields}
\label{pkcs7contenttype}
\end{figure}
\end{center}

CMS defines one protection content type, {\bf ContentInfo},
as the object syntax for documents exchanged between entities. 
ContentInfo encapsulates a single identified content type and the
identified type may provide further encapsulation. A
ContentInfo object contains two fields:
{\tt contentType} (object identifier) and {\tt content}.
CMS defines six {\tt contentType}s: 
data, signed-data, enveloped-data, digested-data, encrypted-data, 
and authenticated-data.  Additional content types can be defined 
outside the CMS document. The type of content can be
determined uniquely by {\tt contentType}. Figure \ref{pkcs7contenttype}
lists the value types in the {\tt content} field for each CMS defined 
content type.
In Figure \ref{pkcs7contenttype},  digestAlgorithms is collection 
of message digest algorithm identifiers. encapContentInfo is the 
signed content, consisting of a content
type identifier and the content itself.
signedAttrs, unsignedAttrs, unprotectedAttrs, authAttrs, and unauthAttrs
are sets of Attribute objects. An Attribute object
is a sequence of two fields: attrType (object identifier) and
attrValues (set of values).

\section{PKCS \#8: Private-Key Information Syntax Standard}
\label{pkcs8}
The security of the public key cryptosystem is entirely dependent
on the protection of the private keys. Generally, the private keys
are encrypted with password and stored in some storage medium.
It is important to have a standard to store private keys
so that one can move private keys from one system to another system
without any trouble.
PKCS \#8 v1.2 describes a syntax for private-key information,
which includes a private key for some public-key algorithm and 
a set of attributes, and a syntax for encrypted private-key information. 
A password-based encryption algorithm 
(e.g., one of those described in PKCS \#5) could be used to encrypt 
the private-key information.

Two objects PrivateKeyInfo and 
EncryptedPrivateKeyInfo are defined in this standard. 
A PrivateKeyInfo object
contains the fields: version, privateKeyAlgorithm, privateKey,
and attributes (optional), where privateKeyAlgorithm
is the identifier of the private key algorithm, privateKey
is the octet string representing the private key, and 
attributes is a collection of attributes that are encrypted
along with the private key.
An EncryptedPrivateKeyInfo object contains two fields:
encryptionAlgorithm and encryptedData,
where encryptionAlgorithm identifies the algorithm under 
which the private-key information is encrypted, and 
encryptedData is the octet string representing the result of 
encrypting the private-key information.

In practice, the PrivateKeyInfo object is BER encoded into an
octet string, which is encrypted with the secret key to give 
the encryptedData field of the EncryptedPrivateKeyInfo object.

\section{PKCS \#9: Selected Object Classes and Attribute Types}
\label{pkcs9}
In order to support PKCS-defined attributes (e.g., to store
PKCS attributes in a directory service) in 
directory systems based on LDAP and the X.500 family of protocols,
PKCS \#9 v2.0 defines two auxiliary object classes, {pkcsEntity} 
and {naturalPerson}. PKCS attributes could be packaged into these
two object classes and be exported to other environments
such as LDAP directory systems. PKCS \#9 v2.0 also defines 
some new attribute types and matching rules that could be used
in other PKCS standards. For example, it defines challengePassword and
extensionRequest attribute types to be used in PKCS \#10 attribute
field, and it defines some attribute types to 
be used in PKCS \#7 (CMS) signedAttrs, unsignedAttrs, 
unprotectedAttrs, authAttrs, and unauthAttrs fields (see 
Section \ref{pkcs7}).
All ASN.1 object classes, attributes, matching rules 
and types defined in PKCS \#9 v2.0 are exported for use 
in other environments.

The {pkcsEntity} object class is a general-purpose auxiliary
object class that is intended to hold attributes about PKCS-related
entities. A {pkcsEntity} object class contains 
fields:
\begin{center}
pkcsEntity = \begin{tabular}{|c|c|c|}\hline
KIND ({auxiliary} type) & PKCSEntityAttributeSet (optional) & ID \\ \hline
\end{tabular}
\end{center}
The {PKCSEntityAttributeSet} may contain any of the
following attributes:
{pKCS7PDU} (with syntax {ContentInfo}), 
{userPKCS12} (with syntax PFX), 
{pKCS15Token} (PKCS \#15), 
{encryptedPrivateKeyInfo} (PKCS \#8),
and future extensions. These attributes should be used
when the corresponding PKCS data (e.g., CMS signed, or enveloped data;
PKCS \#12 personal identity information data; PKCS \#8 encrypted 
private key data, etc.) 
are stored in a directory service.

The {naturalPerson} object class is a general-purpose auxiliary 
object class that is intended to hold attributes about human beings. 
A {naturalPerson} object class  contains 
fields:
\begin{center}
naturalPerson = \begin{tabular}{|c|c|c|}\hline
KIND ({auxiliary} type) & NaturalPersonAttributeSet (optional) & ID \\ \hline
\end{tabular}
\end{center}
The {NaturalPersonAttributeSet} may contain any of the
following (or future extensions) attributes.
\begin{center}\begin{tabular}{|l|l|l|l|l|}\hline
emailAddress&countryOfCitizenship&countryOfResidence&pseudonym&placeOfBirth
   \\ \hline
serialNumber&unstructuredAddress &unstructuredName  &gender   &dateOfBirth
\\ \hline

\end{tabular}\end{center}

PKCS \#9 also defines two matching rules
{pkcs9CaseIgnoreMatch} and {signingTimeMatch} 
which are used to determine whether two PKCS \#9 
attribute values are the same.
Attribute types defined in  PKCS \#9 that are useful in other
standards are listed in Table \ref{pkcs9attributes}.

\vspace{-0.7cm}
\begin{center}
\begin{table}[htb]
\caption{PKCS \#9 Attribute types for use in other standards}
\label{pkcs9attributes}
\begin{center}
\begin{tabular}{|l|p{10cm}|}\hline
{\bf Standard Name} & {\bf Attribute types}\\ \hline
PKCS \#7 and CMS &  contentType, messageDigest, signingTime, 
sequenceNumber, randomNonce, and counterSignature 
(with syntax {SignerInfo})\\ \hline
PKCS \#10 & {challengePassword} (with syntax {DirectoryString}) and
{extensionRequest} (imported from 
ISO/IEC 9594-8 (1997))\\ \hline
PKCS \#12 and \#15 & (user) {friendlyName} and
{localKeyId}\\ \hline
\end{tabular}
\end{center}
\end{table}
\end{center}

\section{PKCS \#10: Certification Request Syntax Standard}
\label{pkcs10}
PKCS \#10 v1.7 specifies syntax for certificate request. When
one entity wants to get a public key certificate, the entity constructs
a certificate request and sends it a certification authority, 
which transforms the request into an X.509 public-key certificate.
A certification authority fulfills the request by authenticating 
the requesting entity and verifying the entity's signature, and, 
if the request is valid, constructing an X.509 certificate from 
the distinguished name and public key, the issuer name, and the 
certification authority's choice of serial number, validity 
period, and signature algorithm. If the certification request 
contains any PKCS \#9 attributes, the certification authority 
may also use the values in these attributes as well as other 
information known to the certification authority to construct 
X.509 certificate extensions. PKCS \#10 does not specify the forms that
the certification authority returns the new certificate.
A certificate request is constructed with the following steps:
\begin{enumerate}
\item Construct a {CertificationRequestInfo} object containing fields:
{version, subject, subjectPKInfo}, and {attributes},
where {subject} contains the entity's distinguished name and
{subjectPKInfo} contains the entity's public key.
Some attribute types that might be useful here are defined in PKCS \#9. 
An example is the challengePassword attribute, which specifies a 
password by which the entity may request certificate revocation. 
Another example is information to appear in X.509 certificate extensions.
\item Sign the {CertificationRequestInfo} object with 
the subject entity's private key.
\item Construct a {CertificationRequest} object containing
fields: {CertificationRequestInfo}, {signatureAlgorithm},
and {signature}, where  {signatureAlgorithm} contains the signature 
algorithm identifier, and  {signature} contains the entity's 
signature.
\end{enumerate}

\section{PKCS \#11: Cryptographic Token Interface Standard}
\label{pkcs11}
PKCS \#11 v2.20 specifies an application programming interface (API), 
called ``Cryptoki'', to devices which hold cryptographic information 
and perform cryptographic functions.  Cryptoki, pronounced ``crypto-key'' 
and short for ``cryptographic token interface'', follows a simple 
object-based approach, addressing the goals of technology independence 
(any kind of device) and resource sharing (multiple applications 
accessing multiple devices), presenting to applications a common, 
logical view of the device called a ``cryptographic token''.
Cryptoki was intended from the beginning to be an interface between 
applications and all kinds of portable cryptographic devices, 
such as those based on smart cards, PCMCIA cards, and smart diskettes.  
The primary goal of Cryptoki was a lower-level programming interface 
that abstracts the details of the devices, and presents to the application 
a common model of the cryptographic device, called a ``cryptographic token'' 
(or simply ``token'').

PKCS \#11 v2.20 specifies the data types and functions available to 
an application requiring cryptographic services using the ANSI C (1990)
programming language.  These data types and functions will typically 
be provided via C header files by the supplier of a Cryptoki library.  
Generic ANSI C header files for Cryptoki are available from the PKCS Web page.

Cryptoki isolates an application from the details of the cryptographic 
device.  The application does not have to change to interface to a 
different type of device or to run in a different environment; thus, 
the application is portable.  

Cryptoki is intended for cryptographic devices associated with a 
single user, so some features that might be included in a general-purpose 
interface are omitted.  For example, Cryptoki does not have a means of 
distinguishing multiple users.  The focus is on a single user's keys 
and perhaps a small number of certificates related to them. Moreover, 
the emphasis is on cryptography.  While the device may perform useful 
non-cryptographic functions, such functions are left to other interfaces.

\begin{center}
\begin{figure}[htb]
\centering{\includegraphics{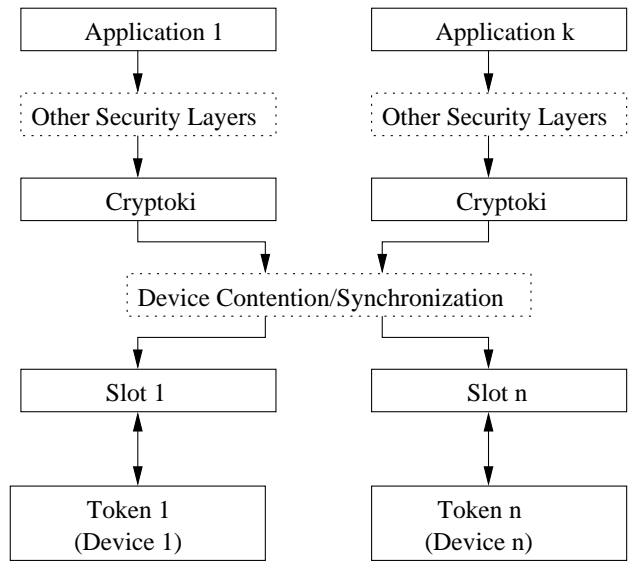}}
\caption{General CryptoKi Model}
\label{pkcs11in}
\end{figure}
\end{center}

Cryptoki is likely to be implemented as a library supporting the 
functions in the interface, and applications will be linked to the library.  
An application may be linked to Cryptoki directly; alternatively, 
Cryptoki can be a so-called shared library (or dynamic link library), 
in which case the application would link the library dynamically.  
The dynamic approach certainly has advantages as new libraries are 
made available, but from a security perspective, there are some drawbacks.  
In particular, if a library is easily replaced, then there is the 
possibility that an attacker can substitute a rogue library that 
intercepts a user's PIN.  From a security perspective, therefore, 
direct linking is generally preferable, although code-signing 
techniques can prevent many of the security risks of dynamic linking.  
In any case, whether the linking is direct or dynamic, the programming 
interface between the application and a Cryptoki library remains the same.
Figure \ref{pkcs11in} describes the general cryptoki model.

Cryptoki defines general data types, objects, and functions.
The general data types include general information data types (e.g.,
CK\_VERSION and CK\_INFO), slot and token types (e.g., CK\_SLOT\_ID),
session types (e.g., CK\_SESSION\_HANDLE), object types 
(e.g., CK\_OBJECT\_CLASS), data types for mechanisms 
(e.g., CK\_MECHANISM\_INFO), function types (e.g., CK\_FUNCTION\_LIST), and 
locking-related types (e.g., CK\_CREATEMUTEX).

Cryptoki's logical view of a token is a device that stores objects
and can perform cryptographic functions.
Cryptoki recognizes three classes of objects, as defined in 
the CK\_OBJECT\_CLASS data type: data, certificates, and keys. 
An object consists of a set of attributes, each of which has a given value. 
A key object stores a cryptographic key. The key may be a public key, 
a private key, or a secret key; each of these types of keys has subtypes 
for use in specific mechanisms (cryptographic algorithms).  
For example, public key objects (object class CKO\_PUBLIC\_KEY) hold 
public keys and contains the following common attributes:  
{\small
\begin{center}
\begin{tabular}{|l|l|l|l|}\hline
CKA\_ID    & CKA\_KEY\_TYPE   & CKA\_DERIVE &CKA\_KEY\_GEN\_MECHANISM\\ \hline
CKA\_WRAP  & CKA\_END\_DATE   & CKA\_LOCAL  &CKA\_KEY\_ALLOWED\_MECHANISM\\ \hline
CKA\_VERIFY& CKA\_SUBJECT     & CKA\_TRUSTED& CKA\_WRAP\_TEMPLATE\\ \hline
CKA\_ENCRYPT & CKA\_START\_DATE & CKA\_CHECK\_VALUE & CKA\_VERIFY\_RECOVER\\ \hline
\end{tabular}
\end{center}
}
According to their lifetime,
objects are classified as ``token objects'' and ``session objects''.
Further classification defines access requirements. ``PIN'' or 
token-dependent methods are required to access ``private token'' while 
no restriction is put on ``public tokens''.

In addition to the PIN protection to private objects on a token,
protection to private keys and secret keys can be given by marking them
as sensitive or unextractable. 
Sensitive keys cannot be revealed in plaintext off the token, and 
unextractable keys cannot be revealed off the token even when 
encrypted (though they can still be used as keys).
It is expected that access to private, sensitive, or unextractable 
objects by means other than Cryptoki (e.g., other programming interfaces, 
or reverse engineering of the device) would be difficult.
Cryptoki does not consider the security of the operating system by 
which the application interfaces to it. For example, since the PIN 
may be passed through the operating system, a rogue 
application on the operating system may be able to obtain the PIN.

Cryptoki provides functions for creating, destroying, and copying 
objects in general, and for obtaining and modifying the values of 
their attributes.  
Objects are always well-formed in Cryptoki. That is, an object always 
contains all required attributes, and the attributes are always 
consistent with the one from the time the object is created.  
This contrasts with some object-based paradigms where an object 
has no attributes other than perhaps a class when it is created, and 
is uninitialized for some time.  In Cryptoki, objects are always initialized.

Cryptoki defines thirteen categories of functions:
        general-purpose functions (4 functions including {\bf C\_Initialize}
and {\bf C\_Finalize}),
        slot and token management functions (9 functions),
        session management functions (8 functions),
        object management functions (9 functions),
        encryption functions (4 functions),
        decryption functions (4 functions),
        message digesting functions (5 functions),
        signing and MACing functions (6 functions),
        functions for verifying signatures and MACs (6 functions),
        dual-purpose cryptographic functions (4 functions),
        key management functions (5 functions),
        random number generation functions (2 functions), and
        parallel function management functions (2 functions).
In addition to these functions, Cryptoki can use 
application-supplied callback functions to notify an application 
of certain events, and can also use application-supplied functions 
to handle mutex objects for safe multi-threaded library access.

Cryptoki has two user types: Security Officer (SO) and normal user.
The function of SO is to initiate a token and to set the PIN for the normal
user. Only the normal user has access to private objects in the token.

A mechanism specifies precisely how a certain cryptographic process 
is to be performed (e.g., a digital signature process or a hashing process). 
Cryptoki defines mechanisms for almost all available cryptographic operations
that are currently used in the industry.

An application in a single address space becomes a 
``Cryptoki application'' when one of its running
threads calls the cryptoki function {\bf C\_Initialize} and it ceases to 
be the ``Cryptoki application'' by calling the 
cryptoki function {\bf C\_Finalize}. Cryptoki has support mechanisms for
multi-threading access.

Cryptoki requires that an application open one or more sessions with 
a token to gain access to the token's objects and functions.  
A session can be a read/write (R/W) session or a read-only (R/O) session.  
R/W and R/O refer to the access to token objects, not to session objects.  
In both session types, an application can create, read, write and 
destroy session objects, and read token objects.  
Table \ref{kievents} lists session events.
\begin{center}
\begin{table}[htb]
\caption{Session events}
\label{kievents}
\begin{center}\begin{tabular}{|l|l|}\hline
Event  & Occurs when...\\ \hline
Log In SO  &     the SO is authenticated to the token.\\ \hline
Log In User &    the normal user is authenticated to the token\\ \hline
Log Out & the application logs out the current user (SO or normal user)
   \\ \hline
Close Session  &  the application closes the session or closes all sessions
   \\ \hline
Device Removed & the device underlying the token has been removed from 
its slot\\ \hline
\end{tabular}\end{center}
\end{table}
\end{center}

Cryptoki header files define a large array of data types. 
Certain packing- and pointer-related aspects of these types are platform- 
and compiler-dependent; these aspects are therefore resolved on a 
platform-by-platform (or compiler-by-compiler) basis outside of 
the Cryptoki header files by means of preprocessor directives.
These directives are described in the Cryptoki also.

\section{PKCS \#12: Personal Information Exchange Syntax Standard}
\label{pkcs12}
PKCS \#12 v1.0 describes a transfer syntax for personal identity 
information, including private keys, certificates, miscellaneous 
secrets, and extensions.  Machines, applications, browsers, Internet 
kiosks, and so on, that support this standard will allow a user to 
import, export, and exercise a single set of personal identity information.
PKCS \#12 can be viewed as building on PKCS \#8 by including 
essential but ancillary identity information along with private 
keys and by instituting higher security through public-key privacy 
and integrity modes.

There are four combinations of {\em privacy modes} and {\em integrity modes}.
The privacy modes use encryption (public-key based or password based) to 
protect personal information from exposure, and the integrity modes 
(public-key digital signature based or password message authentication
code based) protect personal information from tampering. 
For example, in public-key privacy mode, personal information on the source
platform is enveloped using the trusted encryption public key of 
a known destination platform and the envelop is opened using the 
corresponding private-key.

Though all combinations of privacy and integrity modes are permitted,
certain practices should still be avoided. For example, it is unwise
to transport private keys without physical protection when using
password privacy mode. In general, 
it is preferred that the source and destination platforms 
have trusted public/private key pairs usable for 
digital signatures and encryption, respectively.
When trusted public/private key pairs are not available, 
password modes for privacy and integrity could be used.

\begin{center}
\begin{figure}[htb]
\centering{\includegraphics{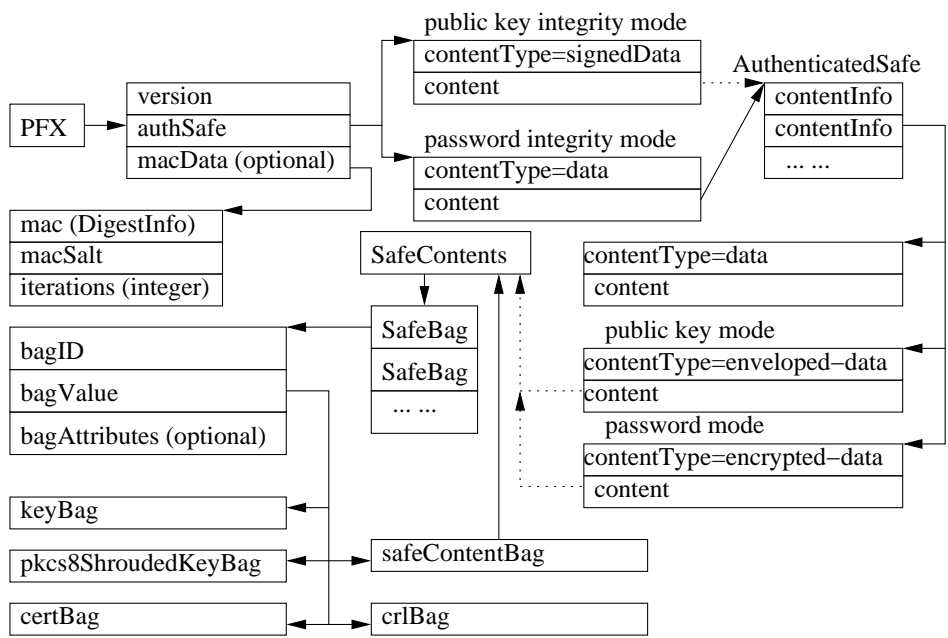}}
\caption{PFX object structure}
\label{pkcs12structure}
\end{figure}
\end{center}

The top-level exchange PDU (Protocol Data Unit) in PKCS \#12
is called PFX. A PFX has three fields: {version, authSafe},
and {macData} (optional), where {authSafe} is a PKCS \#7 
{ContentInfo}. Figure \ref{pkcs12structure} describes the
structure of the PFX object.
It is straightforward to create PFX PDUs from the structure
described in Figure \ref{pkcs12structure}. 
The data wrapped in the PFX could be imported by reversing 
the procedure for creating a PFX. 

\section{PKCS \#15: Cryptographic Token Information Syntax Standard}
\label{pkcs15}
Cryptographic tokens, such as Integrated Circuit Cards (or IC cards) 
are intrinsically secure computing platforms ideally suited to 
providing enhanced security and privacy functionality to applications. 
They can handle authentication information such as digital certificates 
and capabilities, authorizations and cryptographic keys. Furthermore, 
they are capable of providing secure storage and computational facilities 
for sensitive information such as private keys and key fragments.
At the same time, many of these tokens provide an isolated processing 
facility capable of using this information without exposing it within 
the host environment where it is at potential risk from hostile code 
(viruses, Trojan horses, and so on). Unfortunately, the use of these 
tokens for authentication and authorization purposes 
has been hampered by the lack of interoperability. 
First, the industry lacks standards for storing a common 
format of digital credentials (keys, certificates, etc.) on them. 
This has made it difficult to create applications that can work with 
credentials from a variety of technology providers. 
Second, mechanisms to allow multiple applications to 
effectively share digital credentials have not yet reached maturity. 

PKCS \#15 is a standard intended to enable 
interoperability among components running on various platforms 
(platform neutral), to
enable applications to take advantage of products and components 
from multiple manufacturers (vendor neutral), to 
enable the use of advances in technology without rewriting 
application-level software (application neutral), and
to maintain consistency with existing, related standards while expanding upon
 them only where necessary and practical.
As a practical example, the holder of an IC card containing a digital 
certificate should be able to present the card to any application 
running on any host and successfully use the card to present the 
contained certificate to the application. 
As a first step to achieve these objectives, 
PKCS \#15 v1.1 specifies a file and 
directory format for storing security-related information on 
cryptographic tokens. 

\begin{center}
\begin{figure}[htb]
\centering{\includegraphics{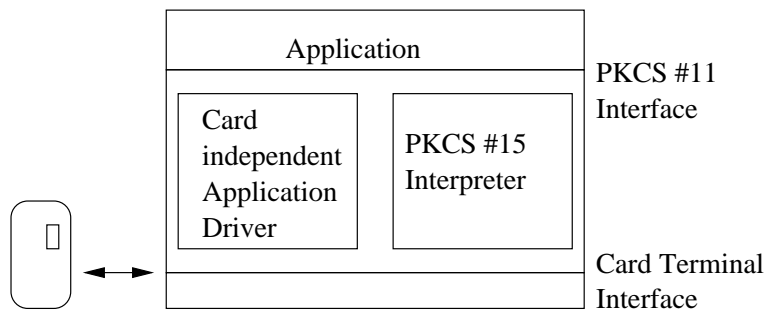}}
\caption{Embedding of a PKCS \#15 interpreter (example)}
\label{pkcs15fig1}
\end{figure}
\end{center}

The PKCS \#15 token information may be read when a token 
is presented, and is used by a PKCS \#15 
interpreter which is part of the software environment, e.g., as shown in
the Figure \ref{pkcs15fig1}.

PKCS \#15 v1.1 defines four general classes of objects: Keys, 
Certificates, Authentication Objects and Data Objects. 
All these object classes have sub-classes, e.g. Private Keys, 
Secret Keys and Public Keys, whose instantiations become 
objects actually stored on cards. 
Objects can be private, meaning that they are protected against 
unauthorized access, or public. In the IC card case, access 
(read, write, etc) to private objects is defined by Authentication 
Objects (which also includes Authentication Procedures). Conditional 
access (from a cardholder's perspective) is achieved with knowledge-based 
or biometric user information. In other cases, such as when PKCS \#15 is 
implemented in software, private objects may be protected against 
unauthorized access by cryptographic means. Public objects are not 
protected from read-access. Whether they are protected against
modifications or not depends on the particular implementation.

\begin{center}
\begin{figure}[htb]
\centering{\includegraphics{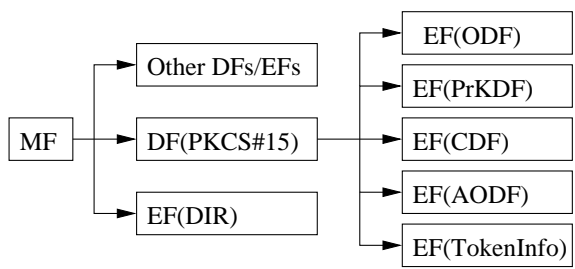}}
\caption{Typical PKCS \#15 Card Layout and Contents of DF(PKCS15)}
\label{pkcs15icf}
\end{figure}
\end{center}

In general, an IC card file format specifies how certain abstract, 
higher level elements such as keys and certificates are to be 
represented in terms of more lower level elements such as IC card files 
and directory structures. A typical IC card supporting 
PKCS \#15 has the file structure layout as in Figure \ref{pkcs15icf}, 
where the following abbreviations are used:
MF (master file), DF($x$) (dedicated file $x$), and 
EF($x$) (elementary file $x$).
PKCS \#15 defines syntax for application directory contents as in Table
\ref{icapp}.
\begin{table}[htb]
\caption{Application Directory Contents}
\label{icapp}
\begin{center}
\begin{tabular}{|l|p{12cm}|}\hline
EF(ODF) & an Object Directory File (ODF)
contains pointers to other EFs (PrKDFs, PuKDFs, 
SKDFs, CDFs, DODFs, and AODFs) \\ \hline
EF(PrKDF) &  a Private Key Directory File (PrKDF) contains 
(references to) private keys\\ \hline
EF(PuKDF) &  a Public Key Directory File (PuKDF) contains 
(references to) public keys\\ \hline
EF(SKDF)  &  a Secret Key Directory File (SKDF) contains 
(references to) secret keys\\ \hline
EF(CDF) & a Certificate Directory File (CDF)  contains 
(references to) certificates\\ \hline
EF(DODF) &  a Data Object Directory File (DODF) is for data objects
other than keys or certificate \\ \hline
EF(AODF) & an Authentication Object Directory File (AODF) is for 
authentication objects such as PINs, passwords, and biometric data \\ \hline
EF(TokenInfo) & a mandatory TokenInfo with transparent structure 
contains generic information about the card (e.g., card serial number,
supported file types, algorithms implemented on the card) and 
it's capabilities\\ \hline
EF(UnusedSpace) & an UnusedSpace file with transparent structure
is used to keep track of unused space in already created elementary files
\\ \hline
 &  other EFs in the PKCS \#15 directory contains the 
actual values of objects (such as private keys, public keys, 
secret keys, certificates and application specific data) referenced 
from within PrKDFs, SKDFs, PuKDFs, CDFs or DODFs  \\ \hline
\end{tabular}
\end{center}
\end{table}
PKCS \#15 compliant IC cards should support direct application
selection as defined in ISO/IEC 7816-4 Section 9 and ISO-IEC 7816-5
Section 6 (the full AID is to be used as parameter for a 
``SELECT FILE'' command). The operating system of the card 
must keep track of the currently selected application and only allow
the commands applicable to that particular application while it is 
selected. The Application Identifier (AID) data element 
consists 12 bytes and its contents is defined in PKCS \#15.

Objects could be created, modified, and removed from the object 
directory file on a card. ASN.1 syntax for these objects have also 
been specified in PKCS \#15.

\section{An Example}
\label{conclusion}
We conclude this chapter with an example application of different
PKCS standards. Assume that we want to implement a smart card 
authentication system based on public key cryptography technology. 
Each user will be issued a smart card containing user's private
key, public key certificate, and other personal information.
Users can authenticate themselves to different computing systems 
(or banking systems) by inserting their
smart cards into card readers attached to these computing systems 
and typing the password (or PIN).

RSA cryptographic primitives specified in PKCS \#1 could be chosen as the
underlying cryptographic mechanisms. First, user Alice needs to 
register herself
to the system to get her smart card. In the registration process,
the system first generates a public-key/private-key pair for Alice.
Using PKCS \#9, the system may create a naturalPerson object or a few
attributes containing Alice's personal information. These information
can then be used to generate a CertificateRequest object according to
PKCS\#10. The system can then send the CertificateRequest object
to the Certificate Authorities (CA) enveloped using 
CMS (PKCS \#7). After the
identity information verification, the CA signs Alice's public key
to generate a certificate for Alice and sends it back to the system.
After receiving Alice's certificate from the CA, the system can now
build a smart card for Alice. Using Alice's password (PIN), the system
generates an EncryptedPrivateKeyInfo object for Alice according to 
PKCS \#8 and PKCS \#9 (PKCS \#5 is also used in this procedure). 
PKCS \#12 may then be used
to transfer Alice's encrypted private key and 
personal information from one computer to another
computer (e.g., from a server machine to the smart card making machine).
Using the dedicated file format DF(PKCS\#15), Alice's encrypted private
key object EncryptedPrivateKeyInfo, certificate, and other personal 
information could be stored on the smart card. The card is now ready
for Alice to use! At the same time, Alice may also get a copy
of these private information on a USB memory stick. These personal information
is stored on the memory stick according to PKCS \#12.

Since all computing systems (e.g., different
platforms from different vendors) support PKCS \#11 API, 
when Alice insert her card into an attached card reader,
applications on these computing systems can communicate smoothly with 
Alice's smart card. In particular, after typing password (PIN),
Alice's smart card can digitally
sign challenges from these computing systems and these computing 
systems can verify Alice's signature using the certificate presented
by Alice's smart card. Thus Alice can authenticate herself to these
systems.

{\bf Acknowledgements}. The author would like to thank anonymous referees
for the constructive comments on improving the presentation of this 
Chapter. The author would also like to thank Dennis Hamilton
(UoL KIT eLearning Division) for some comments on PKCS\#5v2.0.

\newpage
{\Large\bf Glossary}
\begin{enumerate}
\item {\bf  AES} A secret key cipher,  as defined in FIPS PUB 197 (2001)
\item {\bf ASN.1} Abstract Syntax Notation One, as defined in
ISO/IEC 8824-1,2,3,4 (1995)
\item {\bf Attribute} An ASN.1 type that identifies an attribute 
type (by an object identifier) and an associated attribute value
\item {\bf BER} Basic Encoding Rules, as defined in X.690 (1994)
\item {\bf cryptoki}  Short for ``cryptographic token interface''
\item {\bf DES and Triple DES} Secret key ciphers, as defined in 
FIPS PUB 46-3 (1999)
\item {\bf ECC} Elliptic Curve Cryptography
\item {\bf Key derivation function} A function that 
produces a derived key from a base key and other parameters
\item {\bf LDAP} Lightweight Directory Access Protocol, as defined in
Hodges and Morgan (2002)
\item {\bf MAC scheme} A MAC scheme is a cryptographic
scheme consisting of a message tagging operation and a tag checking operation
which is capable of providing data origin authentication and data integrity
\item {\bf MD5} A cryptographic hash function, as defined in Rivest (1992).
MD5 reduces messages of any length to message digests of 128 bits 
\item {\bf OAEP} Optimal Asymmetric Encryption Padding
\item {\bf octet} An octet is a bit string of length 8. 
An octet is represented
by a hexadecimal string of length 2. For example 0x9D represents
the bit string 10011101
\item {\bf octet string} An octet string is an ordered sequence of octets
\item {\bf PDU} Protocol Data Unit, which is a sequence of bits in 
machine-independent format constituting a message in a protocol
\item {\bf personal identity information} Personal information such as 
private keys, certificates, and miscellaneous secrets
\item {\bf PKCS \#11 Token} The logical view of a cryptographic device 
defined by Cryptoki
\item {\bf PKCS \#15 elementary file} 
Set of data units or records that share the same
file identifier, and which cannot be a parent of another file
\item {\bf PKCS \#15 directory (DIR) file} 
Elementary file containing a list of 
applications supported by the card and optional related data elements
\item {\bf SHA-1, SHA-256, SHA-384, and SHA-512} Cryptographic hash function 
functions, as defined in FIPS PUB 180-2, (2002). 
SHA-1 (SHA-256, SHA-384, and SHA-512, respectively)
reduces messages of any length to message digests of 160 bits 
(256 bits, 384 bits, and 512 bits, respectively)
\end{enumerate}

\newpage
\noindent
{\Large\bf References}

Adams, C., and Farrell, S. (1999).
Internet X.509 Public Key Infrastructure
Certificate Management Protocols. IETF RFC 2510.
\url{http://www.ietf.org/}

ANSI C	(1990).  ANSI/ISO 9899: American National Standard for 
Programming Languages - C.

Bellare, M. and Rogaway, P. (1995). Optimal Asymmetric Encryption---How 
to Encrypt with RSA. In A. De Santis, editor, Advances in Cryptology, 
Eurocrypt '94, volume 950 of Lecture Notes in Computer Science, 
pp. 92 - 111. Springer Verlag.

Bellare, M. and Rogaway, P. (1996). The Exact Security of Digital 
Signatures---How to Sign with RSA and Rabin. In U. Maurer, editor, 
Advances in Cryptology,
Eurocrypt '96, volume 1070 of Lecture Notes in Computer Science, 
pp. 399 - 416. Springer Verlag.

Bleichenbacher, D. (1998).
Chosen ciphertext attacks against protocols based on the RSA encryption
standard PKCS \#1. In: {\em Advances in Cryptology '98}, LNCS 1462,
Springer Verlag, 1998.

Diffiem W. and Hellman, M.E. (1976).
New directions in cryptography. IEEE Transactions on Information Theory, 
IT-22:644-654.

FIPS PUB 46-3 (1999). Data Encryption Standard (DES).
US Department of Commerce/National Institute of Standards and Technology.

FIPS PUB 180-2 (2002). Secure Hash Standard (SHS).
US Department of Commerce/National Institute of Standards and Technology.

FIPS PUB 197 (2001). Specification for the Advanced Encryption Standard (AES).
US Department of Commerce/National Institute of Standards and Technology.

Hodges, J. and Morgan, R. (2002). 
Lightweight Directory Access Protocol (v3), IETF RFC 3377.

Housley R. (2002). Cryptographic Message Syntax (CMS), IETF RFC 3369.

IEEE 1363-2000. Standards specification for public key cryptography.
IEEE Press.

IEEE 1363a (2003). Standards specification for public key cryptography: 
additional techniques. Draft.

ISO/IEC 8824-1,2,3,4 (1995).
Information technology - Abstract Syntax Notation One (ASN.1) -
 Specification of basic notation.

ISO/IEC 9594-8 (1997). Information technology - Open Systems 
Interconnection - The Directory: Authentication framework.

Kaliski, B.S. (1993a). An overview of the PKCS standards. RSA Laboratories,
a Division of RSA Data Security, Inc.
\url{http://www.rsasecurity.com/rsalabs/pkcs/}

Kaliski, B.S. (1993b). 
A Layman's Guide to a Subset of ASN.1, BER, and DER.
RSA Laboratories,
a Division of RSA Data Security, Inc.
\url{http://www.rsasecurity.com/rsalabs/pkcs/}

Lenstra Arjen K., and Verheul,  Eric R. (2001).
Selecting Cryptographic Key Sizes.
Journal of Cryptology, 14(4), pp. 255-293.

PKCS (2003). Public-Key Cryptography Standards \#1, \#3, \#5,
\#6, \#7, \#8, \#9, \#10, \#11, \#12, \#15. RSA Laboratories,
a Division of RSA Data Security, Inc.
\url{http://www.rsasecurity.com/rsalabs/pkcs/}

Rivest, R. (1992). The MD5 Message-Digest Algorithm. 
IETF RFC 1321.

Rivest, R., Shamir, A., and Adleman L. (1978). A Method for Obtaining 
Digital Signatures and Public-Key Cryptosystems. Communications of 
the ACM, 21(2), pp. 120-126.

SECG (2000). Standards for Efficient Cryptography Group. Certicom Corp.
\url{http://www.secg.org/}

X.500 (1988). ITU-T Recommendation X.500: The Directory-Overview of 
Concepts, Models and Services.

X.501 (1988). ITU-T Recommendation X.501: The Directory-Models. 

X.509 (2000). ITU-T Recommendation X.509:
The Directory-Public-Key and Attribute Certificate Frameworks. 

X.690 (1994). ITU-T Recommendation X.690:
Information Technology-ASN.1 Encoding Rules: Specification of 
Basic Encoding Rules (BER), Canonical Encoding Rules (CER), 
and Distinguished Encoding Rules (DER).
\end{document}